\documentclass[prd, twocolumn, showpacs, showkeys, floatfix]{revtex4}

\usepackage{graphicx}
\usepackage{amsmath, amsfonts, amssymb, bm}
\begin{document}

\title{Nonlinear Bethe-Heitler pair creation with attosecond laser pulses at the LHC}

\author{Carsten M\"uller}
\affiliation{Max-Planck-Institut f\"ur Kernphysik, Saupfercheckweg 1, 69117 Heidelberg, Germany}

\date{\today}

\begin{abstract}
The creation of lepton pairs ($e^+e^-$ and $\mu^+\mu^-$) via multiphoton absorption in collisions of ultrarelativistic ion beams with ultrashort high-frequency laser pulses is considered. Both the free and the bound-free production channels are addressed, where in the latter case the negatively charged lepton is created in a bound atomic state. It is shown that these nonlinear QED processes are observable when a table-top source of intense xuv or x-ray laser radiation is operated in conjunction with the LHC. We discuss the relative effectiveness of protons versus Pb ions and specify for each pair production channel the most suitable collision system.
\end{abstract}

\keywords{lepton-antilepton pair creation, multiphoton processes, ultrashort laser pulses}
\pacs{12.20.Ds, 13.60.Hb, 32.80.Wr, 34.90.+q }

\maketitle

\section{Introduction}
The Large Hadron Collider (LHC) at CERN (Geneva, Switzerland) is currently starting operation \cite{LHC}. It will reach ion energies up to 7 TeV/nucleon corresponding to proton beams with a Lorentz factor of $\gamma\approx 7000$ and Pb nuclei with $\gamma\approx 3300$. The main physics objectives of this new machine are the search for the Higgs boson and for physics beyond the standard model \cite{Yao}. In this note we describe an application of the LHC for exploring a yet unobserved nonlinear quantum electrodynamic (QED) process, namely lepton-antilepton pair creation in combined nuclear and laser fields. Also other QED processes in this kind of field combination are presently under the active scrutiny of theoreticians, such as photon-fusion or Delbr\"uck scattering \cite{ADP}. This research domain might become interesting for LHC once its major tasks have been accomplished.

The creation of $e^+e^-$ pairs in intense laser fields is encountering a growing interest in recent years \cite{Ringwald}. It has been stimulated by a pioneering experiment at SLAC (Stanford, USA) where $e^+e^-$ pair creation was observed in the collision of a 30 GeV $\gamma$-photon with an optical laser pulse of $10^{18}$\,W/cm$^2$ intensity \cite{SLAC}. The high-energy photon was first produced by Compton backscattering of the same laser beam off a 46 GeV electron beam. Due to the high photon density in the intense laser pulse, the simultaneous absorption of more than one laser photon is possible with a non-negligibly small probability. In the experiment, $n=5$ laser photons of $\hbar\omega_0\approx 2$\,eV combined their energies with the $\gamma$-photon upon the collision to overcome the pair creation threshold: $\gamma + n\omega_0\to e^+e^-$ (nonlinear Breit-Wheeler process). The characteristic feature of intense laser radiation to allow for such multiphoton processes is intensively being studied at lower energies in atomic  physics \cite{Becker}. 

Inspired by the SLAC result, several theoreticians have calculated the production of $e^+e^-$ pairs in head-on collisions of relativistic ion beams with laser radiation (see \cite{Dremin,MVG,Hamlet,Kaminski,Milstein,Kuchiev,Tinsley} and references therein). In this situation the Compton channel is suppressed by the large projectile mass and pair creation occurs predominantly through the nonlinear Bethe-Heitler mechanism where a virtual photon from the Coulomb field combines with $n$ real laser photons to produce the pair:
\begin{eqnarray}
\label{nBH}
\gamma^\ast + n\omega_0 \to e^+e^-\,.
\end{eqnarray}
The counterpropagating-beam geometry enhances the laser frequency to $\omega\approx 2\gamma\omega_0$ in the nuclear rest frame by the relativistic Doppler shift. While the usual Bethe-Heitler process by a single high-energy photon was measured already in the late 1940s \cite{BHexp,Gahn,Erik,Scheid}, its nonlinear generalization (\ref{nBH}) has not been observed yet. The theoretical proposals have been based on the hypothetical combination of two large-scale facilities: (a) either the LHC together with a petawatt-class infrared laser of intensity $I\sim 10^{21}$\,W/cm$^2$ such as the HERCULES laser at the University of Michigan (Ann Arbor, USA) \cite{Yanovsky} or the VULCAN laser at Rutherford Appleton Lab (Didcot, UK) \cite{Vulcan}; (b) or an ordinary but still powerful ion accelerator providing $\gamma\gtrsim 50$ combined with an x-ray free-electron laser (XFEL) of frequency $\hbar\omega_0\approx 10\,$keV; corresponding XFEL beam lines are presently under development at SLAC and DESY (Hamburg, Germany) \cite{XFEL}. These mergings are difficult to realize since in all cases the ion accelerator and laser facility are large in size and located at different places. (Note that the HERA proton accelerator at DESY has been shut down in 2007, unfortunately.)

In the present paper we show that a solution to this obstacle is offered by a new development in laser science. Via high-harmonic generation from laser-irradiated atomic gas jets it has recently become possible in various laboratories to produce intense trains of attosecond laser pulses with frequencies in the xuv domain ($\hbar\omega_0\sim 100$\,eV) and focussed intensities of $\sim 10^{14}$\,W/cm$^2$ \cite{atto-reviews,4labs}. Even higher intensities might be reachable with attosecond pulses from plasma surface harmonics \cite{surface}. These ultra-short pulses hold promising prospects for time-resolved spectroscopy of the electron dynamics in atoms and molecules. Here, however, we propose an unusal application of the attosecond technology in high-energy physics. Since the corresponding experimental devices are of table-top size it is conceivable to bring them to LHC. The Doppler-shifted laser frequency in the ion frame would be sufficiently large to generate $e^+e^-$ pairs by few-photon absorption. We present corresponding production rates for proton and Pb impact, which indicate experimental feasibility of the process. We also briefly discuss $\mu^+\mu^-$ pair creation by utilizing envisaged table-top x-ray devices.

\section{Electron-positron pair creation}
From a theoretical point of view the available attosecond pulse trains (APTs) are weak in the sense that the Lorentz-invariant intensity parameter $\xi=eE_0/m\omega_0$ is much smaller than unity. Here, $E_0$ denotes the peak electric field strength of the linearly polarized laser beam, $-e$ and $m$ are the electron charge and mass, respectively, and we employ natural units ($\hbar=c=1$). For the available APTs its value is $\xi\sim 10^{-4}$. In this regime of laser-matter coupling a process involving $n$ photons could be calculated within $n$-th order of perturbation theory in the photon field. It is, however, more convenient to work within the framework of laser-dressed QED employing relativistic Volkov states \cite{LL,strongfieldQED}. The latter are exact analytical solutions of the Dirac equation in the presence of an external plane-wave laser field. They allow to include the interaction of the leptons with the laser field up to all orders. With regard to nonlinear Bethe-Heitler pair creation, the remaining interaction with the nucleus may be treated in lowest-order of perturbation theory. Within this approach, the amplitude taken in the nuclear rest frame reads
\begin{eqnarray}
\label{S}
S = -ie\int d^4x\, \Psi^\dagger_{p_-,s_-}(x) V_{\rm ion}(r)\, \Psi_{p_+,s_+}(x)\,,
\end{eqnarray}
with the static nuclear potential $V_{\rm ion}(r)= Ze/r$, the nuclear charge number $Z$, and the Volkov states $\Psi_{p_\pm, s_\pm}(x)$. The leptons are characterized by their four-momenta $p_\pm$ and spin projections $s_\pm$ outside the laser field. Although the laser field is treated as a classical electromagnetic wave, photons arise from a mode expansion of the oscillatory parts in Eq.\,\eqref{S} which contains multiphoton processes of arbitrary order. The leading-order rate for nonlinear Bethe-Heitler pair creation by absorption of $n$ laser photons scales like $R^{(n)}\sim \xi^{2n}$. Accordingly, the SLAC experiment found a reaction rate $\sim \xi^{10}$. Higher-order corrections stem from additional photon exchange which goes beyond the minimum photon number required from energy conservation. For example, when pair creation is energetically possible by two-photon absorption there will be small additional contributions where more than two photons have been absorbed. These corrections to the leading-order term are suppressed by an additional $\xi^2$ factor. As an alternative to the $S$-matrix approach, the laser-dressed polarization operator can be employed to calculate total pair creation rates \cite{Milstein}.

The lowest-order term of the rate expansion in powers of $\xi^2$ gives the famous result of Bethe and Heitler for pair creation by a single photon. In the high-energy limit $\omega\gg 2m$, the rate can be written in closed-form as
\begin{eqnarray}
\label{BHhigh}
R^{(1)} = \frac{7}{18\pi}(\alpha Z)^2\xi^2\omega
         \left[ \ln\left({2\omega\over m}\right) - {109\over 42} \right],
\end{eqnarray}
referring to the rest frame of the nucleus. Here, $\alpha$ denotes the QED fine-structure constant. The rate \eqref{BHhigh} can be converted into a cross section by dividing out the photon flux $j=\omega m^2\xi^2/8\pi\alpha$. This result is independent of the photon polarization. We note that in multiphoton physics total probabilities are preferably given as reaction rates since the cross section for a multiphoton process $\sigma^{(n)}=R^{(n)}/j$ still involves powers of $\xi^2$ and thus depends on the incoming photon intensity. The next-to-leading order correction term to Eq.\,\eqref{BHhigh} has been obtained in \cite{Milstein}:
\begin{eqnarray}
\label{delta}
\Delta R = -\frac{13}{90\pi}(\alpha Z)^2\xi^4\omega
   \left[ \ln\left({2\omega\over m}\right) - {22\over 13}\ln 2 - {124\over 195}\right]
\end{eqnarray}
and describes processes where either two photons create the pair or a second photon is absorbed by one of the leptons after the creation. The correction \eqref{delta} would have an appreciable effect in stronger laser fields with $\xi\gtrsim 0.1$ where it reaches the percent level. Corresponding xuv intensities are likely to be attainable from plasma harmonics \cite{surface}. Signatures of the additional photon involved would also arise in the energy spectra of the created particles.
In the opposite nonrelativistic limit close to the energetic threshold ($\omega-2m\ll m$) the one-photon Bethe-Heitler rate reads
\begin{eqnarray}
\label{BHlin}
R^{(1)} = \frac{(\alpha Z)^2}{96}\xi^2\omega\left({\omega-2m\over m}\right)^3.
\end{eqnarray}
For frequencies in the range $m<\omega<2m$ the Bethe-Heitler process can only proceed nonlinearly via two-photon absorption. The result corresponding to Eq.\,\eqref{BHlin} now reads
\begin{eqnarray}
\label{BH2}
R^{(2)} = \frac{(\alpha Z)^2}{64}\xi^4\omega\left({\omega-m\over m}\right)^2
\end{eqnarray}
for a linearly polarized laser wave \cite{Milstein}. It is interesting that the frequency scaling becomes $\sim (\omega-m)^4$ for circular laser polarization. Polarization dependence is a characteristic feature of nonlinear processes, in general.

Let us consider the collision of an intense APT ($\omega_0=100$\,eV, $\xi=1.4\times 10^{-4}$) with the LHC ion beams. For proton impact with $\gamma = 7000$ an $e^+e^-$ pair is produced by one-photon absorption since $\omega=1.4$\,MeV. By numerical calculations based on Eq.\,\eqref{S} we obtain a corresponding lab-frame rate of $R_{\rm lab}^{(1)}\approx 5.79\times 10^2$\,s$^{-1}$ which is in reasonable agreement with the threshold formula \eqref{BHlin}. The latter slightly overestimates the rate by a factor of 2 since the frequency value lies above its range of applicability. Note that the rate in the lab-frame is reduced by a factor $\gamma^{-1}$ because of relativistic time dilation with respect to the ion-frame. When Pb projectiles with $\gamma = 3300$ are used instead, two photons are needed to overcome the energy barrier. The higher-order in $\xi^2$ leads to a smaller total rate of $R_{\rm lab}^{(2)}\approx 2.53\times 10^{-2}$\,s$^{-1}$. Coulomb corrections to the first-order treatment of the nuclear field in Eq.\,\eqref{S} could slightly modify this value. For comparison we note that a proton at the same speed leads to a rate of $R_{\rm lab}^{(2)}\approx 3.76\times 10^{-6}$\,s$^{-1}$. These numbers agree with Eq.\,\eqref{BH2}, again within a factor of 2. Our results are summarized in Table I. The rates can be transformed into total yields by taking the laser pulse length and repetition rate as well as the projectile beam density into account. We assume that in the experiment a single LHC ion bunch containing $N_{\rm ion}\approx 10^{11}$ particles is used. It has a transverse radius of about $\rho_{\rm ion}\approx 16\,\mu$m and circulates with a revolution frequency of $f_{\rm ion}\approx 11$\,kHz \cite{Yao}. An APT of 30\,fs total duration is supposed, consisting of 25 single attosecond pulses with a duration of 300\,as each. The effective time when the field is present in the APT thus amounts to $\tau=7.5$\,fs. Note here that the individual attosecond bursts are separated by half a period of the driving optical laser. 
The train repetition rate can be synchronized with the circulating ion beam, i.e. $f_{\rm APT}=f_{\rm ion}$. The typical diameter of an APT is on the order of 10\,$\mu$m so that we may assume perfect overlap with the ion beam. The number of pair creation events per unit of time is determined by ${\dot N}_{\rm ev}=R_{\rm lab} N_{\rm ion} f_{\rm ion} \tau/2$, with the factor of 1/2 arising from the relative beam velocity. We obtain 0.1 nonlinear Bethe-Heitler pair creation events per second via two-photon absorption from the APT colliding with the LHC Pb beam. This event rate seems to render experimental observation feasible. For comparison we note that about 2400 $e^+e^-$ pairs per second are produced through the ordinary (linear) Bethe-Heitler process when the 7 TeV proton beam is used instead.

We point out that the duration of the single attosecond spikes in the APT amounts to 300\,as/$2\gamma\sim 30$\,zs in the projectile frame \cite{Matveev,Grobe}. This value approaches the natural QED time scale of $1/m\sim 1$\,zs = 10$^{-21}$\,s. The combination of ultrarelativistic particle beams with ultrashort laser pulses might therefore allow for pump-probe experiments on the structure of the physical vacuum when a first zeptosecond spike prepares a certain electron-positron state which is analyzed by a subsequent spike. The pump-probe technique is very successfully applied in atomic physics for time-resolved measurements of processes occuring on the fs and sub-fs scale \cite{atto-reviews}.

\begin{widetext}
\begin{table}[h]
\caption{Lepton pair creation rates in collisions of ultrarelativistic proton and Pb beams with xuv or x-ray laser pulses. The rates were obtained by numerical calculations based on Eq.\,\eqref{S} and refer to a single projectile ion. The xuv (x-ray) laser intensity is $1.78\times 10^{14}$\,W/cm$^2$ ($2.54\times 10^{22}$\,W/cm$^2$). The laser fields are assumed monochromatic and linearly polarized.}
\label{t.1}
\begin{center}
\begin{tabular}{lcccccc}
\hline
 produced pair
 &\ \ projectile\ \
 &\ \ \ $\gamma$ \ \ \
 &\ \ \ $\hbar\omega_0$ [keV] \ \ \
 &\ \ $\hbar\omega$ [MeV] \ \
 &\ photon order \
 &\ lab-frame rate [s$^{-1}$]\  \cr\hline\hline
 free $e^+e^-$        & p  & 7000  & 0.1 & 1.4  & 1 & $5.79\times 10^2$ \cr
                      & p  & 3300  & 0.1 & 0.66 & 2 & $3.76\times 10^{-6}$ \cr 
                      & Pb & 3300  & 0.1 & 0.66 & 2 & $2.53\times 10^{-2}$ \cr\hline 
 bound-free $e^+e^-$  & p  & 7000  & 0.1 & 1.4  & 1 & $6.09\times 10^{-3}$ \cr 
                      & Pb & 3300  & 0.1 & 0.66 & 2 & $4.08\times 10^{-1}$ \cr\hline
 free $\mu^+\mu^-$    & p  & 7000  & 12  & 168  & 2 & $8.25\times 10^{-5}$ \cr
                      & p  & 3300  & 12  &  79  & 3 & $7.77\times 10^{-15}$\cr
                      & Pb & 3300  & 12  &  79  & 3 & $6.59\times 10^{-13}$ \cr\hline
\end{tabular}
\end{center}
\end{table}
\end{widetext}

\section{Bound-free electron-positron pair creation}
A variant of the Bethe-Heitler process (1) is represented by bound-free $e^+e^-$ pair creation \cite{Matveev,Grobe2,boundfree}. Here the $e^-$ is created not as a free particle but in a bound atomic state of the projectile ion. The corresponding process has been observed in relativistic heavy-ion collisions, for example at CERN-SPS \cite{RHI,RHIC}. Bound-free pair creation by a high-energy photon in a nuclear field was, in principle, already calculated by Sauter in the early 1930s \cite{Sauter,Agger}. Its multiphoton generalization in laser fields can be calculated in a way similar to Eq.\,\eqref{S}, with the electronic state given by a bound Coulomb-Dirac wave function. When $n=2$ laser photons participate in the process, the production rate approximately scales like \cite{boundfree}
\begin{eqnarray}
\label{bf}
R_{\rm bf}^{(2)} \sim Z^5 \xi^4 \left(\frac{\omega-\omega_{\rm min}}{m}\right)^\kappa.
\end{eqnarray}
Here, $\omega_{\rm min}$ denotes the threshold photon frequency determined by $n\omega_{\rm min}=2m-\varepsilon_a$, with the atomic binding energy $\varepsilon_a$. The exponent $\kappa$ depends again on the laser polarization: $\kappa\approx0.5$ for linear polarization and $\kappa\approx1$ for circular polarization. The $Z$ scaling is typical for an electron capture process \cite{RHIC}. 
When an LHC proton collides with an xuv pulse, bound-free pair creation proceeds by single photon absorption (Sauter limit) with a rate of $R_{\rm bf}^{(1)}\approx 6.09\times 10^{-3}$\,s$^{-1}$, as shown in Table I. For Pb impact the absorption of two photons is required so that an additional $\xi^2\sim10^{-8}$ factor appears in the total rate. The reduction by this factor, however, is overcompensated by the strong $Z$ dependence in Eq.\,(\ref{bf}). As a result, the two-photon rate from Pb is larger than the one-photon rate from protons and amounts to $R_{\rm bf}^{(2)}\approx 4.08\times 10^{-1}$\,s$^{-1}$. In both cases the electron is assumed to be captured to the $1s_{1/2}$ ground state which gives the major contribution to the process due to its broad momentum distribution. The contribution from higher atomic levels slightly enhances the rate by $\sim 15\%$ \cite{Agger,boundfree}. The rate for nonlinear bound-free pair creation by Pb impact is larger by an order of magnitude than the corresponding creation rate of free pairs. This is due to the stronger $Z$ scaling in combination with the weaker frequency dependence of the bound-free channel [cp. Eqs.\,\eqref{BH2} and \eqref{bf}]. The total number of events, taking the spatio-temporal beam parameters into account, is 1.6\,s$^{-1}$ which should be accessible to experiment. It is interesting to note that antihydrogen atoms could be produced in collisions of APTs with antiprotons, with the rate $R_{\rm bf}^{(1)}$ given above. The corresponding process in relativistic antiproton-ion collisions was realized a decade ago at the CERN-LEAR facility \cite{antihydrogen}.

\section{Muon pair creation}
With x-ray rather than xuv frequencies also nonlinear $\mu^+\mu^-$ pair creation could be realized at LHC \cite{muons}. Corresponding table-top sources of intense APTs with frequencies up to the keV domain are anticipated by surface harmonics in the relativistic regime of laser-plasma interaction \cite{surface}. Moreover, apart from the large-scale XFEL facilities presently under development, there are also efforts to build table-top XFEL devices \cite{Gruner}. According to optimistic estimates, these might reach comparable operation parameters due to very high quality of the driving electron beam. $\mu^+\mu^-$ production requires higher laser frequencies because of the large muon mass $m_\mu\approx 207m$ and the rate scaling $\sim \xi_\mu^{2n}$, where $\xi_\mu=\xi m/m_\mu$ and $n\sim m_\mu/\omega$. In contrast to $e^+e^-$ creation, the finite size of the projectile nucleus has to be taken into account here since the typical momentum transfer $q\sim m_\mu$ is of the order of the inverse nuclear radius. As usual, this introduces the nuclear form factor $F=F(q^2)$ into the reaction rate by virtue of \cite{Tsai}
\begin{eqnarray}
\label{muon}
R_\mu = R_0[Z^2 F^2 + Z(1-F^2)].
\end{eqnarray}
The term $\sim Z^2$ accounts for the elastic channel where the nucleus remains in its ground state and the $Z$ protons act coherently; the term $\sim Z$ describes the inelastic channel involving nuclear excitation and incoherent proton action. $R_0$ denotes the reaction rate for a pointlike proton; for the case of two-photon muon creation close to threshold, $R_0$ is given by Eq.\,\eqref{BH2} with the replacements $m\to m_\mu$ and $\xi\to \xi_\mu$. For simplicity we employ in our calculations a spherically symmetric Gaussian charge distribution, whose free parameter is adjusted to reproduce the measured rms charge radius. We assume an x-ray beam of frequency $\omega_0 = 12$\,keV and intensity parameter $\xi_\mu = 6.8\times 10^{-5}$ \cite{XFEL}. This corresponds to an intensity of order $10^{22}$\,W/cm$^2$ which can be achieved by improved x-ray focussing techniques. When colliding with the LHC proton beam, muon pair creation is possible via two-photon absorption and occurs with a rate of $R_\mu^{(2)}\approx 8.25\times 10^{-5}$\,s$^{-1}$ (see Table I). This result has already been reported in \cite{muons}. We note that a similar rate can be expected for $\pi^+\pi^-$ creation. The LHC Pb beam produces muons with much lower efficiency since three photons are needed to bridge the energy gap, which suppresses the process by a factor $\xi_\mu^2\sim 10^{-8}$. Besides, the production mainly proceeds through the inelastic channel because of the large nuclear size of Pb, so that its charge enhances the rate by a factor $Z$ only. As a consequence, Coulomb corrections are of minor importance here \cite{RHIC}. The resulting rate is $R_\mu^{(3)}\approx 6.59\times 10^{-13}$\,s$^{-1}$ which seems hardly observable. Experimental studies of $\mu^+\mu^-$ production should therefore rely on the LHC proton beam. The total event rates are reduced by the fact that the beams overlap only partially since a sub-$\mu$m waist size of the x-ray pulse is required to obtain sufficiently high intensity. As a background process, also $e^+e^-$ pairs are produced by single-photon absorption in the nuclear field. The rate is given by Eq.\,\eqref{BHhigh} and amounts to $\sim Z^2\times 10^{11}\,{\rm s}^{-1}$ in the lab frame. We emphasize that this rather strong background process does not deplete the x-ray beam: assuming a pulse duration of 100\,fs and a focal radius of 100\,nm, the latter contains about 10$^{15}$ photons, whereas only $\sim Z^2\times 10^{-2}$ $e^+e^-$ pairs are generated per ion.

Finally, we point out that $e^+e^-$ pair production might also become feasible within an all-optical setup where a relativistic proton beam is generated by laser wakefield acceleration instead of the LHC. In the so-called laser-piston regime, proton energies corresponding to $\gamma\approx 50$ have been predicted \cite{Esirkepov}. According to the simulations, each proton bunch contains $\sim 10^{12}$ particles and has a transverse radius of about $5\,\text{$\mu$m}$. When combined with an ultrashort x-ray pulse produced from plasma harmonics ($\omega_0\approx 5\!-\!10$\,keV), the energy threshold for $e^+e^-$ pair creation would be overcome by two-photon absorption leading to a total rate of $R_{\rm lab}^{(2)}\approx 5$\,s$^{-1}$.

\section{Conclusion}
In conclusion, lepton pair creation via the nonlinear Bethe-Heitler process in relativistic laser-ion collisions has been considerd. A feasible setup for the first observation of this process could be obtained by combining two cutting-edge technologies which are usually considered unrelated, namely an attosecond xuv photon source with CERN-LHC. In contrast to earlier proposals this combination is, in principle, readily available since such laser pulses are produced by table-top devices. In conjunction with the LHC Pb beam, event rates approaching ${\dot N}_{\rm ev}\sim 1$\,s$^{-1}$ could be achieved for the creation of free or bound-free $e^+e^-$ pairs via absorption of two real photons. By envisaged compact x-ray devices and the LHC proton beam also $\mu^+\mu^-$ pairs could be produced in this manner, though at lower rates \cite{annihilation}. Nonlinear laser polarization effects could be tested and contributions from next-to-leading order terms revealed. The zeptosecond duration of the pulses in the ion frame holds prospects for time-resolved pump-probe studies on the QED vacuum.

\section*{Acknowledgements}
The author thanks K. Z. Hatsagortsyan and C. H. Keitel for helpful discussions and carefully reading the manuscript.


\begin{thebibliography}{99}

\bibitem{LHC} {\it The LHC Design Report}, edited by O. Br\"uning, et al., CERN
Report No. 2004-003 (http://lhc.web.cern.ch).

\bibitem{Yao} W.-M. Yao, et al., J. Phys. G 33 (2006) 1.

\bibitem{ADP} A. Di Piazza, K. Z. Hatsagortsyan, C. H. Keitel, Phys. Rev. Lett. 100 (2008) 010403; A. Di Piazza, A. I. Milstein, Phys. Rev. A 77 (2008) 042102.

\bibitem{Ringwald}
A. Ringwald, Phys. Lett. B 510 (2001) 107;
M. Marklund, P. Shukla, Rev. Mod. Phys. 78 (2006) 591;
Y. I. Salamin, et al., Phys. Rep. 427 (2006) 41.

\bibitem{SLAC} D. Burke, et al., Phys. Rev. Lett. 79 (1997) 1626.

\bibitem{Becker} 
D. B. Milo\v{s}evi\'c, et al., J. Phys. B 39 (2006) R203.

\bibitem{Dremin} I. M. Dremin, Pis'ma Zh. Eksp. Teor. Fiz. 76 (2002) 185 [JETP Lett.
76 (2002) 151].

\bibitem{MVG} C. M\"uller, A. B. Voitkiv, N. Gr\"un, Phys. Rev. A 67 (2003) 063407;
Phys. Rev. A 70 (2004) 023412.

\bibitem{Hamlet} H. K. Avetissian, et al., Nucl. Instrum. Meth. Phys. Res. A 507 (2003) 582.

\bibitem{Kaminski} 
P. Sieczka, et al., Phys. Rev. A 73 (2006) 053409;
K. Krajewska, J. Z. Kaminski, Laser Phys. 18 (2008) 185.

\bibitem{Milstein} A. I. Milstein, et al., Phys. Rev. A 73 (2006) 062106.

\bibitem{Kuchiev} M. Yu. Kuchiev, Phys. Rev. Lett. 99 (2007) 130404.

\bibitem{Tinsley}
The corresponding process of $e^+e^-$ pair creation by a high-energy neutrino colliding with an intense laser beam has been calculated by T. M. Tinsley, Phys. Rev. D 71 (2005) 073010. The production proceeds through an intermediate $Z^0$ boson and naturally occurs with a tiny rate.

\bibitem{BHexp}
G. P. Adams, Phys. Rev. 74 (1948) 1707;
W. Heitler, \textit{The Quantum Theory of Radiation} (Clarendon Press, Oxford, 1954).

\bibitem{Gahn}
Bethe-Heitler pair creation in a high-$Z$ target via bremsstrahlung photons from laser-accelerated electrons has been observed in C. Gahn, et al., Appl. Phys. Lett. 77 (2000) 2662; Phys. Plasmas 9 (2002) 987.

\bibitem{Erik}
Laser-assisted Bethe-Heitler pair creation by a single high-energy photon has been treated recently in E. L\"otstedt, U. D. Jentschura, C. H. Keitel, Phys. Rev. Lett., in press.

\bibitem{Scheid}
Pair creation accompanying high-energy photoionization has recently been calculated by 
E. G. Drukarev, A. I. Mikhailov, I. A. Mikhailov, Kh. Yu. Rakhimov, W. Scheid,
Phys. Rev. A 75 (2007) 032717.

\bibitem{Yanovsky}
V. Yanovsky, et al., Opt. Express 16 (2008) 2109;\\
http://www.engin.umich.edu/research/cuos/.

\bibitem{Vulcan}
http://www.clf.rl.ac.uk/Facilities/vulcan/laser.htm.

\bibitem{XFEL} 
L. F. DiMauro, et al., J. Phys. Conf. Ser. 88 (2007) 012058; M. Altarelli, et al., Technical Design Report of the European XFEL, DESY 2006-097 (http://www.xfel.net).

\bibitem{atto-reviews}
P. Agostini, L. F. DiMauro, Rep. Prog. Phys. 67 (2004) 813;
M. Drescher, F. Krausz, J. Phys. B 38 (2005) S727;
A. Scrinzi et al., J. Phys. B 39 (2006) R1;
D. Charalambidis, et al., New J. Phys. 10 (2008) 025018;
C. Winterfeldt, C. Spielmann, G. Gerber, Rev. Mod. Phys. 80 (2008) 117.

\bibitem{4labs}
P. Tzallas, et al., Nature 426 (2003) 267;
Y. Mairesse, et al., Science 302 (2003) 1540;
K. J. Schafer, et al., Phys. Rev. Lett. 92 (2004) 023003;
G. Sansone, et al., Science 314 (2006) 443;
P. B. Corkum, F. Krausz, Nat. Phys. 3 (2007) 381.

\bibitem{surface} A. Pukhov, Nature Phys. 2 (2006) 439; G. D. Tsakiris, et al., New J. Phys. 8 (2006) 19; N. M. Naumova, et al., New J. Phys. 10 (2008) 025022.

\bibitem{LL} V. B. Berestetskii, E. M. Lifshitz, L. P. Pitaevskii, {\it Relativistic
Quantum Theory} (Pergamon Press, London, 1971).

\bibitem{strongfieldQED}
H. R. Reiss, J. Math. Phys. 3 (1962) 59;
H. Mitter, Acta Phys. Austr. Suppl. 14 (1975) 397;
V. I. Ritus, J. Sov. Laser Res. 6 (1985) 497;
A. I. Nikishov, J. Sov. Laser Res. 6 (1985) 619;
W. Becker, Laser Part. Beams 9 (1991) 603;
T. Heinzl, O. Schr\"oder, J. Phys. A 39 (2006) 11623.

\bibitem{Grobe}
P. Krekora, Q. Su, R. Grobe, Phys. Rev. Lett. 92 (2004) 040406;
J. Mod. Opt. 52 (2005) 489.

\bibitem{Matveev}
V. I. Matveev, E. S. Gusarevich, I. N. Pashev, J. Exp. Theor. Phys. 100 (2005) 1043.

\bibitem{boundfree} C. M\"uller, A. B. Voitkiv, N. Gr\"un, Phys. Rev. Lett.
91 (2003) 223601; C. Deneke, C. M\"uller, Phys. Rev. A 78 (2008) 033431.

\bibitem{Grobe2}
P. Krekora, K. Cooley, Q. Su, R. Grobe, Phys. Rev. Lett. 95 (2005) 070403.

\bibitem{RHI}
H. F. Krause, et al., Phys. Rev. Lett. 80 (1998) 1190.

\bibitem{RHIC} 
J. Eichler, Phys. Rep. 193 (1990) 165;
K. Rumrich, K. Momberger, G. Soff, W. Greiner, N. Gr\"un, W. Scheid, Phys. Rev. Lett. 66 (1991) 2613;
G. Baur, K. Hencken, D. Trautmann, Phys. Rep. 453 (2007) 1.

\bibitem{Sauter}
The results on photoionization by F. Sauter, Ann. Physik 11 (1931) 454, can be transformed into bound-free pair creation rates via the usual crossing symmetry.

\bibitem{Agger}
C. K. Agger, A. H. S{\o}rensen, Phys. Rev. A 55 (1997) 402.

\bibitem{antihydrogen}
G. Baur et al., Phys. Lett. B 368 (1996) 251.

\bibitem{muons}
C. M\"uller, C. Deneke, C. H. Keitel, Phys. Rev. Lett. 101 (2008) 060402.

\bibitem{Gruner}
F. Gr\"uner, et al., Appl. Phys. B 86 (2007) 431.

\bibitem{Tsai} 
A. Alberigi-Quaranta, et al., Phys. Rev. Lett. 9 (1962) 226; 
Y.-S. Tsai, Rev. Mod. Phys. 46 (1974) 815.

\bibitem{Esirkepov} T. Esirkepov, et al., Phys. Rev. Lett. 92 (2004) 175003.

\bibitem{annihilation}
Muon pair creation via $e^+e^-$ annihilation in intense low-frequency photon fields has been calculated by C. M\"uller, K. Z. Hatsagortsyan, C. H. Keitel, Phys. Lett. B 659 (2008) 209; see also Ref.~\cite{Kuchiev} and M. H. Thoma, arXiv:0801.0956.

\end{thebibliography}
\end{document}